\documentclass[11pt]{article}

\usepackage[T1]{fontenc}
\usepackage{newtxtext,newtxmath}
\usepackage[margin=1in]{geometry}
\usepackage{graphicx}
\usepackage{amsmath}
\usepackage{booktabs}
\usepackage{float}
\usepackage{array}
\usepackage[hidelinks]{hyperref}
\usepackage{url}
\makeatletter
\renewcommand\normalsize{%
  \@setfontsize\normalsize{11}{13.6}%
  \abovedisplayskip 11\p@ \@plus3\p@ \@minus6\p@
  \abovedisplayshortskip \z@ \@plus3\p@
  \belowdisplayshortskip 6.5\p@ \@plus3.5\p@ \@minus3\p@
  \belowdisplayskip \abovedisplayskip
  \let\@listi\@listI}
\makeatother
\normalsize
\providecommand{\doi}[1]{\url{https://doi.org/#1}}

\newcommand{\fld}[1]{\texttt{#1}}

\title{\fontsize{18}{22}\selectfont\textbf{What Students Actually Ask: Demand Structure and Automation Potential in a Hybrid Support System}}
\author{%
\large Jinal Gupta$^{1,*}$, Pavani Ayinampudi$^{1}$, Aditya B.M.V.$^{1}$, Prakash Hegade$^{1}$,\\
\large Rohit Sharma$^{2}$, Sakshi Sharma$^{2}$, Meenakshi V$^{2}$, S.R.S. Iyengar$^{2}$\\[8pt]
\normalsize $^{1}$ANNAM.AI, Indian Institute of Technology Ropar, Rupnagar, Punjab, India\\
\small\texttt{\{jinalbirla, pavania.harvard2025, adityabmv, prakash.hegade\}@gmail.com}\\[4pt]
\normalsize $^{2}$Indian Institute of Technology Ropar, Rupnagar, Punjab, India\\
\small\texttt{\{rohit.24csz0014, sakshi.23csz0006,}\\
\small\texttt{meenakshi.19csz0013, sudarshan\}@iitrpr.ac.in}\\[6pt]
\small $^{*}$Corresponding author: \texttt{jinalbirla@gmail.com}}
\date{}

\begin{document}
\maketitle
\begin{abstract}
Large online programmes receive heavy volumes of queries during onboarding, at
a scale that grows faster than the number of staff available to answer them.
This paper reports a study of an AI integrated query resolution platform that
spreads incoming queries across four routes: an AI based assistant, answers from
fellow participants, a curated corpus of frequently asked questions, and
escalation to administrators. Over nine weeks, from 2 May to 6 July 2026, the
system handled 4,093 queries raised by 1,434 participants. Nearly every query reached a recorded resolution, and one query in five closed
within an hour.
Reuse of 114 corpus entries absorbed 21.3\% of the volume, participants resolved
a further 12.2\% on their own, and 132 participants answered questions for one
another at a median of 9 to 15 minutes, showing that peer answering, where it
occurred, was fast and broadly shared across the cohort. Classifying the
query text shows that demand was narrow rather than varied. A single process
step, the submission of a certificate and the offer letter that follows it,
accounts for 56.8\% of corpus mediated resolutions, and at least 20.4\% of
queries concern the progress of a pending submission rather than a request for
information, a class the assistant served only 1.3\% of the time, since a stored
answer cannot report an individual's current status. Only 8.0\% of the queries
handled by a person duplicated content already in the corpus, which indicates
that the knowledge base was already well used. Most direct administrative closures occur in synchronous bulk events, a pattern
that shapes how the records should be read.

\end{abstract}


\section{Introduction}

Large scale online programmes experience substantial volumes of learner queries
during the onboarding phase. A student joining an internship programme meets a
question at almost every step: what to do next, in what order, what happens if a
document is late, which message will arrive once a form is filed. Each step
depends on the one before it, so a small early uncertainty can hold a
participant in place until it is resolved. As enrolment grows, the demand for
support outpaces the growth in available staffing, and answering every student
individually becomes difficult to sustain.

Our aim was to resolve these queries with a short turnaround, without assigning
a dedicated mentor to each student, which would not scale. We first curated a
corpus of frequently asked questions recording, in detail, the procedure for
each step, whom to contact, and the messages a student could expect after each
action. The corpus was revised daily, so a change to a rule or a class timing
was reflected the same day, and recurring situations were added as they were
observed. A rich corpus alone proved insufficient, however, because students did
not always locate the entry that answered their question, so we built a pipeline
that directed each query to the relevant entry and, where a query spanned
several, composed a single answer drawing on each.

We then engaged the participants themselves. Rather than route every query
through a small staff team, we let any student answer a peer's query, by pointing
to an existing entry or drafting a new one. A peer answer did not reach the asker
directly: it passed to an administrator, who could accept, revise, or return it,
and an accepted answer went to a senior administrator for final approval before
release. This kept quality under control while opening participation to the wider
cohort, and let the corpus grow from the questions students actually raised.

This paper reports what happened when the system ran, using nine weeks of its
operational records. We ask what participants actually asked, since the right
response depends on the nature of the demand, and how far the peer layer helped
resolve it. The first question matters because varied conceptual questions call
for answering capacity, repeated procedural questions call for wider corpus
coverage, and enquiries about the progress of a pending action call for neither,
since the answer changes from day to day and belongs to one participant alone.
These profiles cannot be told apart without reading the query text, and they
point to different design choices. Our contribution is an empirical account of
support demand across 4,093 queries from 1,434 participants, an estimate of how
much of that demand the existing corpus already covered, and two properties of
the resolution record that qualify what such records can evidence. Throughout we
attend to the peer layer, since a central promise of such a system is that
participants can resolve one another's queries quickly and at scale.

Section 2 reviews related work and identifies the gap. Section 3 sets out the
study context, research questions, and design. Section 4 describes data
collection and analysis. Section 5 reports and discusses the results with their
limitations. Section 6 concludes.

\section{Background Study}

The system studied here draws on three lines of work: how learners seek help,
how automated and crowd based answering has met that need, and how the
resulting records can be read.

Help seeking has long been treated as a behaviour worth studying rather than a
neutral signal of need. Aleven et al. \cite{ref_aleven}, reviewing two decades
of research on help seeking in intelligent tutoring systems, report that
learners often under use or misuse the help available to them, and that
improving help seeking does not reliably improve outcomes. That work, together
with research relating help seeking to broader engagement \cite{ref_lam},
concerns help with academic content. The demand we examine is operational,
covering eligibility, documentation and scheduling, and this difference in kind
has received little systematic attention.

Where the volume of such questions is high, automated answering has been a
natural response. Goel and Polepeddi \cite{ref_goel} describe Jill Watson, a virtual teaching
assistant that answered routine forum questions in a large online course. Later
work extended this with generative models grounded in course material
\cite{ref_taneja,ref_kakar}, and related studies examined how the scope of the
corpus shapes what such an agent can answer \cite{ref_eicher} and how agents
fill other instructional roles \cite{ref_law}. Assistants of this
kind have since been applied to administrative as well as academic enquiries
\cite{ref_dinh,ref_ghazouani}, including support for international students
\cite{ref_wang}. Systematic reviews report that chatbots in higher
education serve mainly teaching, service and wellbeing functions and are
typically evaluated on satisfaction or accuracy
\cite{ref_labadze,ref_kuhail,ref_okonkwo,ref_segovia}; the composition of the
demand they receive is rarely reported, a gap this paper addresses. Since an
assistant restricted to verified content cannot answer beyond it, grounding
responses in a curated corpus guards against unsupported generation
\cite{ref_ji,ref_karpukhin}, and it is the design adopted here.

When automated answering reaches its limit, the participants themselves become a
resource, which connects our peer layer to learnersourcing. Khosravi et al.
\cite{ref_khosravi} set out the design and analytics agenda for systems in which
learners generate content for one another, and a substantial literature
addresses how such contributions are quality assessed at scale and how
participation is sustained \cite{ref_abdi,ref_darvishi,ref_denny,ref_moore,ref_moore25}.
Such assistance has been shown effective regardless of who authored it, whether
teacher help redistributed beyond its class \cite{ref_patikorn} or crowd
authored feedback in programming education \cite{ref_aljumeily}. This work
usually concerns pedagogical artefacts; our peer layer produces operational
answers instead, but the questions of who contributes and how effort is
distributed carry over directly, and our data allow us to report them.

Two further findings inform how we read the results. Anderson et al.
\cite{ref_anderson} show how the value of a contribution in community question
answering shifts from the asker toward an enduring knowledge base, which
presumes diverse questions with reusable answers. Buell and Norton
\cite{ref_buell} show, in a service setting, that making the work behind a
process visible changes how waiting is experienced even when its duration is
unchanged.
Peer support latency has also been measured directly, with a median of nine
minutes on a platform in an unrelated domain \cite{ref_morris}, and forum
studies note that instructor presence has a non monotonic effect on peer
answering \cite{ref_chandrasekaran}. The closest methodological precedent lies outside education, in support ticket
classification. Al-hawari and Barham \cite{ref_alhawari} describe a machine
learning help desk with mature techniques for categorising free text requests.
That work is concerned with routing accuracy, however, rather than with what the
mix of request types implies for service design.

Across these strands, the incoming demand is generally taken as given. This
study sits between them. It characterises the composition of operational support
demand in a deployed hybrid system, and relates that composition to how queries
were dispatched, what each path cost, how much the existing corpus already
covered, and what the peer layer contributed.

\section{Methods and Methodology}

\subsection{Methodology and Study Context}

This is a single site, retrospective deployment study using complete
operational records. The unit of analysis is the individual query, and the
approach is descriptive rather than causal, since no comparison condition exists
and the configuration was fixed throughout the observation window.

The setting is a large scale online internship programme whose participants join
in staggered cohorts and complete a fixed sequence of onboarding steps before
project work: declaring intent, uploading an institutional No Objection
Certificate, receiving an offer letter, confirming dates, and enrolling in
preparatory courses. Each step depends on the previous one, and several require
staff verification, which introduces waiting periods during which a participant
cannot proceed.

The platform routes a raised query first to the AI chatbot, which answers only
from a curated corpus of verified FAQ entries, a deliberate restriction that
prevents unverified procedural guidance \cite{ref_ji}. Queries it cannot serve
pass to the peer layer, where any participant may answer. A peer answer reaches
the asker only after an administrator accepts or revises it and a senior
administrator gives final approval; administrators may also answer directly, and
a scripted facility covers queries a standard response can handle. The authors
operate this system, which gives direct knowledge of the workflow while creating
an interest in how it is described, addressed in Section 5.3.

\subsection{Research Questions}

The study is guided by a main research question, addressed through three
subsidiary questions that build to it.

\begin{description}
\item[RQ.] In a large scale, continuously enrolling online internship
programme, how is support demand composed and dispatched across a hybrid AI,
peer and administrator resolution system, and under what conditions can
operational resolution records be read as a fair account of that demand?
\end{description}

\noindent The subsidiary questions are:

\begin{description}
\item[RQ1.] When queries are classified by what they request rather than by how
they were resolved, what distinct demand types emerge among participants, and
does demand type carry information that resolution path alone cannot?

\item[RQ2.] How is each demand type dispatched across the automated, peer,
administrative and knowledge reuse paths, what does each path cost in time to
closure, and how much of the demand reaching a person was already answerable
from existing verified content?

\item[RQ3.] Where the record diverges from what its fields appear to state, in
attribution, in outcome coding and in queries that resist routine closure, what
accounts for the divergence, and what does this imply for reading such records
as evidence of contribution and of quality?
\end{description}

RQ1 establishes what the demand consists of and whether that composition is
visible from dispatch data alone. RQ2 establishes how the demand was handled,
at what cost, and how much of it could have been handled without human effort.
RQ3 tests the evidential standing of the record itself and therefore qualifies
the answers to the first two.

\subsection{Research Design}

The design combines two analyses over the same corpus. The first classifies
query text to establish what participants asked; the second analyses dispatch
and timing to establish how queries were handled. Categories from the first are
applied to the second, so demand type and resolution path can be cross
tabulated. Classification was performed without reference to resolution
outcome.

\section{Data Collection and Analysis}

\subsection{Data Source}

Prior to data collection, informed consent was obtained from all participating
students regarding the use of their data for research purposes.

The dataset is a complete export of the platform's operational records covering all
queries raised between 2 May and 6 July 2026, comprising 4,093 records from
1,434 distinct participants. Records span two merged schema variants, a legacy
variant with 708 records and a current variant with 3,385, reflecting a
platform migration during the collection period. Each record contains the
raising timestamp, query text, resolution status, resolver type, resolver
identity where recorded, response timestamp, linked FAQ identifier where
knowledge reuse occurred, escalation trigger where applicable, and a derived
acceptance flag. No records were excluded except where stated. Timestamps were
parsed as ISO 8601 UTC without timezone conversion.

\subsection{Classification of Demand}

Categories were derived inductively: the first author read successive random
samples and extracted recurring patterns, iterating until unclassified volume
fell below 25\%. The scheme contains nine categories and is applied by a
priority ordered rule based classifier, assigning each query to its most
specific match. Tuned for precision, patterns were narrowed whenever inspection
revealed false positives, leaving a substantial residual unclassified.
Precision was measured on a stratified sample of 108 labelled queries, twelve
per category. This is an automated classification with measured precision, not
a thematic analysis: boundaries were set by one author, no second coder
participated, and no inter rater statistic \cite{ref_falotico} is available, so
category volumes are reported as lower bounds.

\subsection{Dispatch and Timing Analysis}

Elapsed time was computed between the raising and response timestamps for every
record holding both. Medians and the 75th, 90th and 95th percentiles are
reported by path, since timing distributions are heavily right skewed and the
tail carries the experience most consequential for participants. Records
without a closure timestamp are reported as a censored count.

\subsection{Estimation of Corpus Coverage}

To estimate how much demand reaching a person was already covered by verified
content, query text was compared against the corpus. The export contains FAQ
identifiers but not answer text, so each entry was approximated by the queries
it resolved, and the 871 corpus linked queries stand as a coverage proxy. All
query text was vectorised using TF-IDF over unigrams and bigrams with sublinear
scaling, and cosine similarity was computed between each unlinked query
resolved by a person and its nearest corpus linked neighbour. Control tokens
were excluded from both sides. Results are reported across thresholds from 0.50
to 0.85 with a manually validated sample. TF-IDF captures lexical rather than
semantic correspondence, so the estimate is conservative for paraphrase, and
denser retrieval \cite{ref_karpukhin} would be expected to raise it.

\subsection{Repeat Enquiry and Concentration}

Repeat enquiries were identified where the same participant raised a lexically
similar query, cosine similarity at or above 0.60, within seven days of a
previous query that had already closed. Concentration of demand across
participants is reported as a Lorenz curve and Gini coefficient, and
concentration of reuse across corpus entries by the same statistics.

\section{Results and Discussion}

\subsection{Results}

\paragraph{RQ1: Composition of demand.}
Classification of all 4,093 queries produced the distribution in
Table~\ref{tab5}. Overall precision on the validation
sample was 75.0\%, with per category values given in the same table.
Misclassifications were asymmetric: items belonging to status enquiry were
assigned elsewhere but not the reverse, so its reported proportion is a lower
bound.

\begin{table}[!htb]
\caption{Demand categories from rule based classification ($N = 4{,}093$).}
\label{tab5}
\centering
\begin{tabular}{|l|r|r|r|}
\hline
Category &  n & \% & Validated precision\\
\hline
Status enquiry & 833 & 20.4 & 12/12\\
Procedural information & 790 & 19.3 & 9/12\\
Technical fault & 463 & 11.3 & 9/12\\
Logistics and cohort & 240 & 5.9 & 6/12\\
Eligibility edge case & 213 & 5.2 & 6/12\\
Enrolment intent & 180 & 4.4 & 9/12\\
Points and scoring & 139 & 3.4 & 10/12\\
Record correction & 77 & 1.9 & 8/12\\
Uninterpretable control & 255 & 6.2 & 12/12\\
Unclassified & 903 & 22.1 & n/a\\
\hline
\end{tabular}
\end{table}

Of the 148 queries resolved by the AI assistant, 60 were procedural
information, 42 enrolment intent and 11 status enquiry. The last figure
corresponds to 1.3\% of all status enquiries. Status enquiries closed at a
median of 40.8 hours. Status enquiry accounts for 109 of the 501 self resolved
queries, or 21.8\%, against 20.4\% of the corpus overall.

\paragraph{RQ2a: Dispatch and cost by path.}
Table~\ref{tab:paths} gives volume and time to closure by resolution path.
Scripted and direct administrative action together account for the majority of
volume, while the automated and peer paths handle the smallest shares.
Knowledge reuse covered 871 queries, 21.3\%, drawing on 114 distinct corpus
entries, a mean of 7.6 reuses per entry, with the most reused entry linked to
49 queries.

\begin{table}[!htb]
\caption{Resolution path: volume and time to closure ($N = 4{,}093$). Timing
statistics cover records holding both a raising and a closure timestamp; for
self resolution these number 382 of 501, and the three paths marked with a dash
carry no reported timing.}\label{tab:paths}
\centering
\footnotesize
\begin{tabular}{|l|r|r|r|r|r|r|}
\hline
resolver\_type &  n & \% & p50 & p90 & p95 & \% $>$72h\\
\hline
\fld{bulk\_admin\_script} & 1,294 & 31.6 & 50.7 h & 281.0 & 502.8 & 41.3\\
\fld{human\_admin\_direct} & 1,122 & 27.4 & 47.0 h & 138.9 & 495.5 & 28.6\\
\fld{bulk\_admin\_script\_legacy} & 706 & 17.2 & 25.2 h & 259.6 & 264.5 & 39.5\\
\fld{self\_resolved\_by\_asker} & 501 & 12.2 & 18.6 min & 153.2 & 234.2 & 18.3\\
\fld{peer\_answered\_no\_formal\_approval} & 225 & 5.5 & 15.0 min & 78.8 & 203.2 & 11.1\\
\fld{ai\_chatbot} & 148 & 3.6 & -- & -- & -- & --\\
\fld{peer\_genuine\_approve} & 61 & 1.5 & 9.2 min & 21.5 & 69.2 & 4.9\\
\fld{human\_resolved\_unattributed} & 33 & 0.8 & -- & -- & -- & --\\
\fld{human\_admin\_direct\_legacy} & 2 & $<$0.1 & 108.5 h & 110.1 & 110.3 & 100.0\\
\fld{unresolved\_or\_other} & 1 & $<$0.1 & -- & -- & -- & --\\
\hline
\textit{all closed queries} & 3,792 & & 28.9 h & 242.4 & 304.4 & 32.5\\
\hline
\end{tabular}
\end{table}

Reuse is unevenly spread across entries (Table~\ref{tab4}). Entries 4, 3 and 1,
all concerning certificate submission and offer letter issuance, account for
495 of 871 resolutions, or 56.8\%, and no other cluster exceeds 9.0\%. Across
the 113 well formed entries the Gini coefficient of reuse is 0.586: the ten most
reused absorb 38.0\%, while 30 entries, 26.5\% of the corpus, were used exactly
once. Median reuse per entry is 3 against a mean of 7.7. Entries 6 and 9 contain
a disproportionate share of links with no discernible question in the
associated text.

\begin{table}[!htb]
\caption{Knowledge reuse volume by corpus entry number ($N = 871$). One record
carries a compound identifier naming two entries and is assigned to neither, so
per-entry counts total 870.}\label{tab4}
\centering
\begin{tabular}{|c|l|r|r|}
\hline
Entry \# &  Inferred topic & n & \%\\
\hline
4 & Certificate status and offer letter issuance & 345 & 39.6\\
3 & Phase eligibility and certificate format & 102 & 11.7\\
10 & Points system and course scheduling & 78 & 9.0\\
2 & Examination conflicts and internship timing & 51 & 5.9\\
1 & Certificate template and institutional eligibility & 48 & 5.5\\
12 & Phase and course details, offer letter status & 44 & 5.1\\
6 & Off-campus forms and onboarding tickets & 38 & 4.4\\
5 & Domain assignment and dashboard faults & 33 & 3.8\\
13 & Cohort logistics and breakout sessions & 33 & 3.8\\
9 & Points system mechanics & 31 & 3.6\\
11 & AI assistant access & 24 & 2.8\\
7 & Portal and interview status faults & 18 & 2.1\\
14 & Team formation & 14 & 1.6\\
8 & Communication channel access & 11 & 1.3\\
\hline
\end{tabular}
\end{table}

Time to closure separates sharply by path. Within one hour of raising, 70.3\%
of peer answered queries had closed, against 1.9\% of those answered directly
by an administrator and 14.8\% of those closed by script. Aggregate statistics across all closed queries are given in the final row of
Table~\ref{tab:paths}. A further 301 queries, 7.4\%, carry no closure timestamp
and are censored.

\begin{figure}[!htb]
\centering
\includegraphics[width=0.5\textwidth]{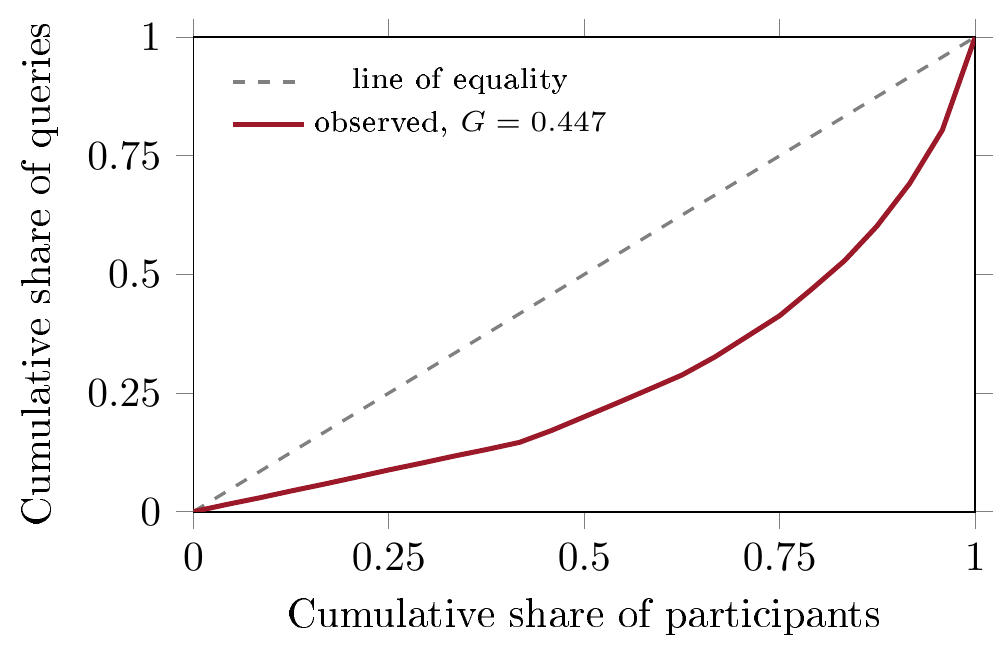}
\caption{Lorenz curve of query volume across the 1{,}434 participants.}
\label{fig:lorenz}
\end{figure}

Across the 1,434 participants, mean demand is 2.85 queries, median 2 and
maximum 36. The most active decile accounts for 34.6\% of queries and the most active fifth for 51.9\%. Exactly one query was
raised by 42.9\% of participants, and 4.0\% raised ten or more. The Lorenz
curve is shown in Fig.~\ref{fig:lorenz}. The least active half of participants
account for 20.0\% of volume and the least active 80\% for 48.1\%.

\paragraph{RQ2b: Corpus coverage of demand reaching a person.}
Excluding control tokens, 2,664 queries resolved by a person carried no corpus
link. At a cosine similarity threshold of 0.70, 8.0\% of these matched an existing
corpus linked query, at a manually validated match precision of 14 of 15.
Table~\ref{tab6} gives sensitivity across thresholds. The estimate falls from
8.0\% to 6.5\% as the threshold rises from 0.70 to 0.85. Of the 213 matched
queries, 45 are status enquiries, leaving 6.3\% of candidates.

\begin{table}[!htb]
\caption{Share of person resolved queries matching existing corpus content, by
similarity threshold ($n = 2{,}664$ candidates).}\label{tab6}
\centering
\begin{tabular}{|l|r|r|r|r|r|r|}
\hline
Threshold & 0.50 & 0.60 & 0.70 & 0.75 & 0.80 & 0.85\\
\hline
Matched queries & 355 & 270 & 213 & 196 & 184 & 173\\
\% of candidates & 13.3 & 10.1 & 8.0 & 7.4 & 6.9 & 6.5\\
\hline
\end{tabular}
\end{table}

Among query texts longer than 25 characters occurring more than once, 109
distinct texts cover 323 queries. Of these, 47 distinct texts covering 183
queries had some instances linked to a corpus entry and other identical
instances resolved without one.

\paragraph{RQ3a: Queries resisting routine closure.}
Of 224 escalation tagged queries, 178 carry the \fld{too\_many\_skips} trigger
and 46 the \fld{too\_many\_ambiguous} trigger, with resolution paths differing
by trigger (Table~\ref{tab3}). Within the ambiguity trigger, 25 of 46 are
uninterpretable control tokens rather than substantive questions. Within the
skips trigger the largest categories are technical fault, status enquiry and
logistics.

\begin{table}[!htb]
\caption{Resolution path by escalation trigger, as percentage within
trigger.}\label{tab3}
\centering
\begin{tabular}{|l|r|r|}
\hline
resolver\_type &  skips ($n=178$) & ambiguous ($n=46$)\\
\hline
\fld{bulk\_admin\_script} & 33.1 & 2.2\\
\fld{self\_resolved\_by\_asker} & 30.3 & 26.1\\
\fld{human\_admin\_direct} & 29.2 & 56.5\\
\fld{human\_resolved\_unattributed} & 7.3 & 15.2\\
\hline
\end{tabular}
\end{table}

The slowest decile comprises 380 queries, all exceeding 242.4 hours. Its
largest categories are routine, status enquiry and procedural information, and
273 of the 380 were closed by scripted bulk action rather than being
intrinsically hard. Repeat enquiries number 22 among the 3,838 interpretable
queries, or 0.6\%.

\begin{figure}[!htb]
\centering
\includegraphics[width=0.7\textwidth]{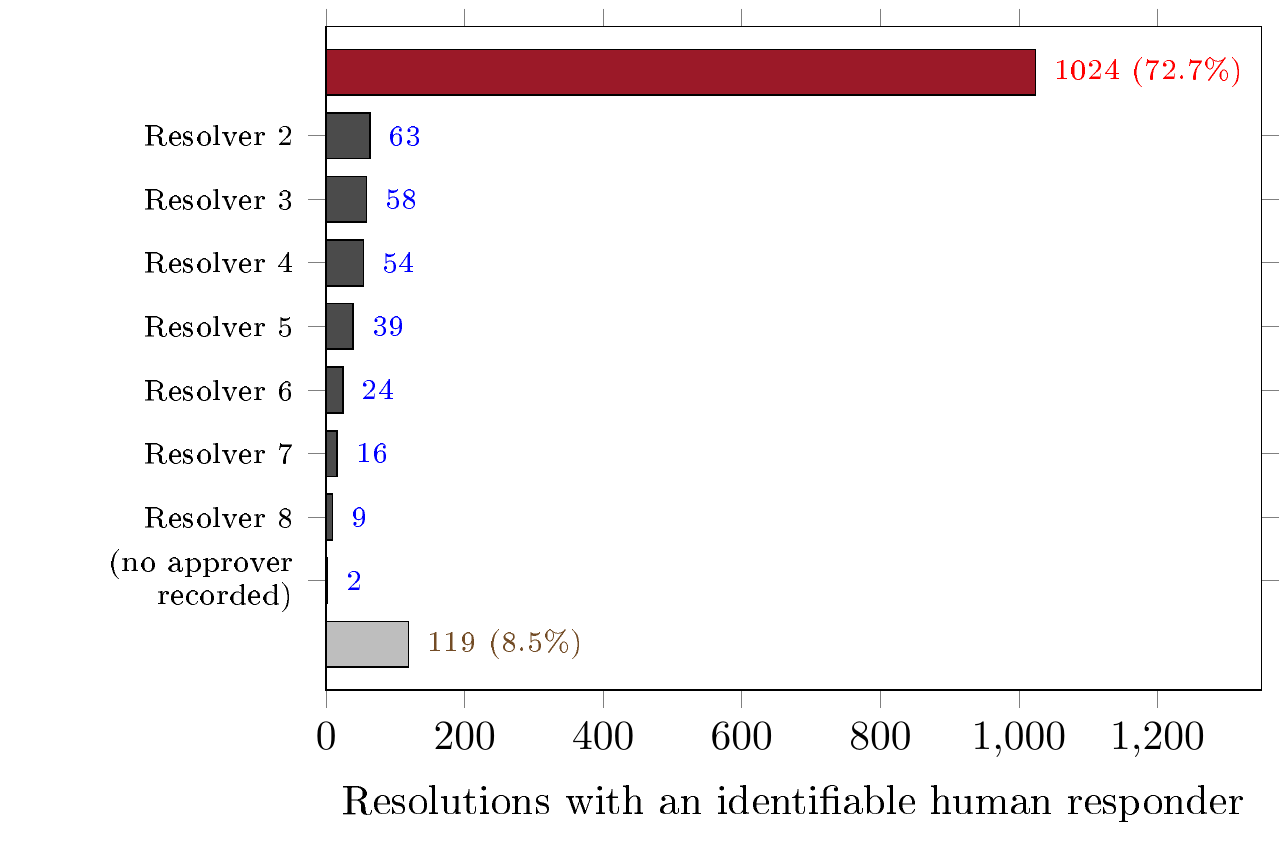}
\caption{Recorded resolver identity across the 1{,}408 resolutions with an
identifiable human responder, identities anonymised.}
\label{fig:resolvers}
\end{figure}

\paragraph{RQ3b: Attribution and outcome coding.}
Among the 1,408 resolutions with an identifiable human responder, one
administrator accounts for 1,024, or 72.7\%, with no other individual exceeding
4.5\%. Of that administrator's 1,024 records, 1,021, or 99.7\%, carry the
\fld{human\_admin\_direct} path, and the administrator accounts for 91.0\% of
all 1,122 records on that path. The full distribution appears in
Fig.~\ref{fig:resolvers}.

Peer typed records carry a peer identifier in 61 of 61
\fld{peer\_genuine\_approve} and 224 of 225
\fld{peer\_answered\_no\_formal\_approval} records, while the resolver email
field holds the approving staff member; only three of the administrator's 1,024
records are peer typed. Peer resolutions number 285, authored by 132
participants of whom 101 also raised queries, at a Gini of 0.414 across authors.
Of the 119 records lacking a resolver email, 118 carry a peer identifier
without a recorded approver.

The \fld{accepted\_derived} field is true for every resolution path except self
resolution and one unresolved record, both of which are uniformly false. This
constitutes a perfect partition on whether the resolver differed from the
asker.

\subsection{Discussion}

Research on help seeking treats a query as a request for knowledge the asker
lacks, and evaluates support systems on whether that knowledge is supplied
\cite{ref_aleven}. RQ1 indicates that this model describes much of the demand observed. A further
class, amounting to at least a fifth, follows a different pattern that dispatch
data alone does not reveal, since status enquiries appear on every resolution
path. A status enquiry asks where a
submission has reached, and its answer is specific to one participant and
changes from day to day. Three observations mark this class as distinct. The
retrieval constrained assistant served 1.3\% of it while handling procedural
questions readily. Status enquiries closed at a median of 40.8 hours although
each is individually simple to answer. They are also more frequent among self
resolutions than in the corpus overall, which is consistent with the awaited
event occurring in the interim. This pattern also accounts for the 1.3\%
figure, which is lower than might be expected of a first line assistant. A
corpus grounded agent serves the demand its corpus covers \cite{ref_eicher},
and a corpus can hold procedure but not a participant's present status, so the
figure marks the boundary of the design working as intended. Administrative
assistants of the same family \cite{ref_dinh,ref_ghazouani} operate within the
same boundary, and reviews of chatbots in higher education document adoption
across teaching, service and wellbeing functions
\cite{ref_labadze,ref_kuhail,ref_okonkwo}. Demand composition is a complementary
question, and this study supplies it for one operational setting.

Work on learnersourcing and knowledge base construction has established how
content should be generated and its quality assessed
\cite{ref_khosravi,ref_abdi,ref_darvishi,ref_denny}, and proceeds on the
reasonable expectation that better content improves outcomes. RQ2 suggests that
this deployment was already close to that ceiling, since 8.0\% of queries
reaching a person duplicated existing content, falling to 6.3\% once status
enquiries are set aside. Gains from further corpus expansion would therefore be
modest here, which is itself a useful result for a team deciding where to
invest. The 183 queries with identical text routed along different paths suggest that
the available margin lies more in dispatch consistency than in coverage. This
is a routing question of the kind the ticket classification literature addresses
\cite{ref_alhawari}. Since the lexical method used here detects
paraphrase conservatively, denser retrieval \cite{ref_karpukhin} would be
expected to raise the estimate, and 8.0\% is best read as a floor.

The same analysis describes what the corpus accumulates. Anderson et al.
\cite{ref_anderson} show how community question answering builds an enduring
knowledge base whose value depends on diverse questions with reusable answers.
The pattern here differs: a Gini of 0.586 across entries, with a quarter used
once and one step accounting for 56.8\% of reuse, indicates concentration on a
single high frequency step rather than breadth. The opportunity lies upstream,
in the step itself, rather than in documenting it further.

Evidence for what an upstream change might achieve comes from service
operations. Buell and Norton \cite{ref_buell} show that making the work behind a
process visible alters how waiting is experienced even when its duration is
unchanged; applying this to educational support demand is, to our knowledge,
new. The status class matches that situation: a participant who can see a
certificate is received and queued for checking has the information the query
would have sought. With demand spread widely across the cohort, at a Gini of
0.447 with 42.9\% raising a single query, visibility into process state offers a
more direct route than additional answering capacity.

These readings rest on the record meaning what its fields suggest, the concern
of RQ3. Because peer answers pass through approval before closure, attribution
might have named the approver rather than the author. It does not: 99.7\% of the
administrator's records are direct answers and peer identity is held separately.
The peer layer is modest in volume yet broadly shared, at a Gini of 0.414 across
132 authors, supplying the contributor distribution that studies of crowd
sourced assistance \cite{ref_patikorn,ref_aljumeily} establish to be effective.
Three properties of the record warrant care on reuse. The resolver email field
denotes the author on administrative paths but the approver on peer paths, and
\fld{accepted\_derived} encodes whether the resolver differed from the asker
rather than acceptance. The third concerns timing. Because most administrative closures occur in
synchronous bulk events, a raw concentration figure conflates answering with
queue clearing, and a raw latency conflates routine turnaround with periodic
backlog clearing. Separating the two distinguishes the routine median of 6.1
hours from the 28.9 hour aggregate, and this separation is visible only in the
distribution of closure timestamps, not in any single field. Establishing workflow semantics before interpreting
contributor distributions is thus a useful step for studies on operational
support records, including in the ticket classification setting
\cite{ref_alhawari}.

\subsection{Limitations}

Five constraints bound these findings. The data cover a single programme across
nine weeks of onboarding, with volume peaking in week 3 and tapering after, so
they describe an onboarding period rather than a steady state, and we claim no
generality across programmes. The demand categories were defined by one author
with a rule based classifier of measured precision; with no second coder there
is no inter rater statistic \cite{ref_falotico}, so the volumes are lower bounds
and the two least precise categories are indicative only. The coverage estimate
compares query text against a proxy corpus built from corpus linked queries,
since the export holds no answer text, so it shows that comparable content
existed rather than that the assistant would have answered correctly. No finding
evaluates resolution quality, since the acceptance field encodes resolver
identity and repeat enquiry, at 0.6\%, is too rare to stand in for it. Finally,
the authors operate the system, which gives both workflow knowledge an outside
analyst would be unlikely to recover and an interest in how it is described; we
address the latter by reporting the RQ3 measurement checks in full, including
those that did not support the reading the field names invite.

\section{Conclusion}

This paper reported a deployment study of an AI integrated query resolution
platform serving a large online internship programme, covering nine weeks and
4,093 queries from 1,434 participants. Nearly every query reached a recorded
resolution. Reuse of 114 corpus entries handled 21.3\% of the volume,
participants resolved a further 12.2\% on their own, and 132 answered questions
for one another at a median of 9 to 15 minutes, so the peer layer, where used,
was both quick and widely shared. Repeat enquiry was rare, at 0.6\%, and only
8.0\% of queries a person handled duplicated existing corpus content, which
suggests the knowledge base was well used.

The make-up of the demand explains the rest of the load. A single process step,
the submission of a certificate and the offer letter that follows, accounts for
56.8\% of corpus reuse, and at least 20.4\% of queries ask about the progress of
a pending submission rather than for information, which the assistant could
serve only 1.3\% of the time, since a stored answer cannot report a student's
current status. Direct administrative answering was concentrated in one
administrator who handled 91.0\% of it, a figure best read alongside the way
closures were recorded (Section 5.2); peer answering was smaller in volume but
widely shared.

Three directions follow. Making the state of a pending process visible to
participants would remove the largest single source of volume without anyone
answering more questions. Recording authorship and approval in separate fields
on every path, as the peer paths already do, would make contribution measurable
system wide. And repeating this study across further cohorts would show how far
the demand profile carries over.

\section*{Acknowledgements}
The authors thank the VicharanaShala Lab for Education Design (VLED), Indian Institute of Technology Ropar, for the opportunity to carry out this research within its internship program, for access to the anonymised operational data, and for supporting the research reported here. We also thank the program's operations team and the participants whose queries made this study possible.

\section*{Disclosure of Interests}
The authors have no competing interests to declare that are relevant to the content of this article.

The authors used a generative AI assistant to help restructure and edit drafts of the text and to format the manuscript. The authors designed the study, checked all analyses, results, and references, and take full responsibility for the content.

\bibliographystyle{splncs04}
\bibliography{refs}

\end{document}